\documentclass[12pt]{article}
\usepackage{graphicx}
\usepackage{amsfonts}
\usepackage{amssymb}
\usepackage{epsf}
\newcommand{\nin}{\noindent}
\newcommand{\be}{\begin{equation}}
\newcommand{\ee}{\end{equation}}
\newcommand{\bea}{\begin{eqnarray}}
\newcommand{\eea}{\end{eqnarray}}
\newcommand{\br}{\hskip .25cm/\hskip -.25cm}

\newcommand{\nn}{\nonumber\\}

\newcommand{\ol}{\overline}

\begin{document}

\nin KCL-PH-TH/2012-{\bf 20}

\begin{center}
{\Large{\bf Spontaneous symmetry breaking and\\ linear effective potentials}}

\vspace{1cm}

J. Alexandre 

King's College London, Department of Physics, WC2R 2LS, UK

jean.alexandre@kcl.ac.uk

\vspace{2cm}

{\bf Abstract}

\end{center}

\nin The convexity of a scalar effective potential is a well known property, and, in the situation of 
spontaneous symmetry breaking, leads to the so-called
Maxwell construction, characterised by a flat effective potential between the minima of the bare potential. 
Simple derivations are given here, which show how linear effective potentials arise
from non-trivial saddles points which dominate the partition function, for a self-interacting scalar 
field and for a Yukawa model.

\vspace{1cm}

\section{Introduction}

The convexity of the effective potential for a scalar theory has been known for a long time \cite{Symanzik},
and is a purely non-perturbative effect, which cannot be reproduced by a naive loop expansion. 
Convexity is a consequence of the definition of the effective potential in terms of a Legendre transform.
In addition, the effective potential exhibits a flat part, and the corresponding Maxwell construction 
is a consequence of non-trivial saddle points in the path integral which defines the quantum theory, 
when the bare potential features spontaneous symmetry breaking (SSB). The aim of this article is to give 
a simple analytical argument for the occurrence of a linear effective potential, based on a semi-classical description.\\ 
A whole PhD thesis has been written on the topic \cite{Kessel}, where many aspects are studied, and a detailed literature
review is given.
A simple argument for the convexity can be found in \cite{Haymaker}, and the role of two competing non-trivial saddle points
for the flattening of the effective potential was proposed in \cite{Fujimoto}.
Lattice approaches
are given in \cite{Callaway} and \cite{Mukaida}, with an analytical argument based on a large $N$ expansion, and where finite size effects
are also discussed. Convexity of the effective potential and its relevance to finite temperature phase transitions is studied in \cite{Rivers}, 
and a treatment of convexity in Quantum Mechanics can be found in \cite{BB}. 
Also, interesting approaches to convexity have been studied, using exact Wilsonian renormalization group, and they are
shortly reviewed here, in subsection 2.3.\\ 
We note that an original realization of the flattening of the potential is given by the spinodal inflation \cite{Holman}. 
In this work, the authors decompose the inflaton as the sum of its space average $\phi$ and the standard deviation $\sigma$ of its fluctuations.
The resulting model is similar to a hybrid inflation scenario, and it is shown that the flattening of the potential seen by $\phi$,
as time increases (or $\sigma$ increases), is a consequence of the instability of modes with super-horizon wave-length.\\
Finally, the spinodal instability also appears in relativistic nuclear collisions \cite{Randrup}, where both quark-gluon plasma and hadrons 
can coexist (spinodal decomposition).
More generally, the flattening of the effective potential is related to the instability of the mono-phasic vacuum, and leads to the
coexistence of two phases, or bubbles of different vacuua.\\
In the next section, general arguments for convexity are reviewed, and a derivation for the Maxwell construction is proposed,
in the case of a self-interacting real scalar field.
Section 3 describes similar steps when the scalar field is coupled to a fermion. It is shown that the presence of a fermion condensate and 
a non-trivial vacuum expectation value (v.e.v) for the scalar field are self-consistent, and lead to a linear potential between the minima of
the bare potential.

\section{Self-interacting real scalar field}

We first review the general arguments leading to the convexity of the effective potential, and then propose a derivation
for the linearity of the effective potential between the minima of the bare potential.

\subsection{Convexity of the effective potential}

We work her in Euclidean metric, which makes the argument of convexity more straightforward. It has been noted, though, that ambiguities 
can arise when using the Minkowski metric \cite{Argyres}.\\
Given a bare action $S[\phi]$, the partition function $Z$ and the connected graph generating functional $W$ are defined as
\be
Z[j]=\int{\cal D}[\phi]\exp\left(-S[\phi]-\int j\phi\right)\equiv\exp\left( -W[j]\right)~,
\ee
and the classical field is 
\be\label{phic}
\phi_c\equiv\frac{\delta W}{\delta j}=\frac{1}{Z}\int{\cal D}[\phi]~\phi~\exp\left(-S[\phi]-\int j\phi\right)
=\left<\phi\right>~.
\ee
$\phi_c$ is a functional of the source $j$, which does not correspond to a physical source, but is rather an intermediate variable,
to parametrize the system, and which will eventually be replaced by the classical field.
A key inequality is obtained by taking the second derivative of $W$ 
\be\label{1stkey}
\frac{\delta^2W}{\delta j\delta j}=-\left<\phi\phi\right>+\left<\phi\right>\left<\phi\right>~\leq0~.
\ee
The proper graph generating functional $\Gamma$, defined as the Legendre transform of $W$,
is obtained by inverting the relation between the source and the classical field, such that $j$ has now to be read as a 
functional of $\phi_c$, and
\be\label{Gamma}
\Gamma[\phi_c]=W[j]-\int j\phi_c ~.
\ee
The effective action $\Gamma$ contains all the quantum corrections which dress the system.\\
The equation of motion of the classical field is obtained by noting that
\be
\frac{\delta\Gamma}{\delta\phi_c}=\int\frac{\delta W}{\delta j}\frac{\delta j}{\delta\phi_c}-j-\int\phi_c\frac{\delta j}{\phi_c}~,
\ee
and using the definition (\ref{phic}) of the classical field, to find
\be
\frac{\delta\Gamma}{\delta\phi_c}=-j~,
\ee
where we remind that $j$ has to be understood as a functional of $\phi_c$.
An important relation is then
\be
\frac{\delta^2\Gamma}{\delta\phi_c\delta\phi_c}=-\frac{\delta j}{\delta\phi_c}=-\left( \frac{\delta\phi_c}{\delta j}\right)^{-1}= 
-\left( \frac{\delta^2W}{\delta j\delta j}\right)^{-1}~,
\ee
which, together with the inequality (\ref{1stkey}), leads to 
\be
\frac{\delta^2\Gamma}{\delta\phi_c\delta\phi_c}~\geq0~.
\ee
The effective potential is the momentum independent part of $\Gamma$, and is therefore a convex function of the classical field.

\subsection{Maxwell construction}

As a consequence of convexity, 
quantum effects must erase all possible concave contribution in the bare potential. We now show how this
happens, taking into account the non-trivial saddle point configurations which appear when the bare potential feature SSB.\\
We assume that $\phi_1>0$ and $\phi_2<0$ are non-vanishing constant configurations for which the bare action $S$ has the local
minima $S_1=S[\phi_1]$ and $S_2=S[\phi_2]$.
The bare action might contain a physical source,
such that we do not necessarily have $\phi_2=-\phi_1$ and $S_1=S_2$.\\
We do not take into account the kink configuration, since the latter is stable in 1+1 dimensions only, if no other field is present 
\cite{Jackiw}, and we consider here two space dimensions at least (a recent work on the quantization of the 1+1 
dimensional kink can be found in \cite{Rajantie}).
In addition, the kink has a finite action $S_{kink}$, and its contribution exp($-S_{kink}$)
to the partition function 
is much smaller than the contribution exp($-S_{1,2}$) of constant saddle points, with negative actions $S_{1,2}$,
which are proportional to the volume.\\ 
If one neglects quantum fluctuations, the partition function is then dominated by the two saddle points $\phi_1,\phi_2$
\be\label{Z}
Z\simeq \exp\left(-S_1-\int j\phi_1\right)+\exp\left(-S_2-\int j\phi_2\right)~,
\ee
and the classical field is
\be\label{phicsp}
\phi_c\simeq\frac{\phi_1\exp(-S_1-\int j\phi_1)+\phi_2\exp(-S_2-\int j\phi_2)}{\exp(-S_1-\int j\phi_1)+\exp(-S_2-\int j\phi_2)}~,
\ee
which represents a weighted average of the two configurations $\phi_1,\phi_2$. This shows that the approximation (\ref{Z})
can describe the region $\phi_2\le\phi_c\le\phi_1$ only: the limits of the classical field (\ref{phicsp}) are 
\bea
\lim_{j\to+\infty}\phi_c&=&\phi_2\nn
\lim_{j\to-\infty}\phi_c&=&\phi_1~.
\eea
Outside the interval $[\phi_2,\phi_1]$, the effective action is equal to the bare action, since we neglect
quantum fluctuations. In this approximation, the effective action is, for $\phi_2\le\phi_c\le\phi_1$,
\be\label{Gammasp}
\Gamma[\phi_c]=-\ln\left[ \exp\left(-S_1-\int j\phi_1\right)+\exp\left(-S_2-\int j\phi_2\right)\right] -\int j\phi_c~,
\ee
and, as expected, has the following limits
\bea
\Gamma[\phi_2]=\lim_{j\to+\infty}\Gamma[\phi_c]&=&S[\phi_2]\nn
\Gamma[\phi_1]=\lim_{j\to-\infty}\Gamma[\phi_c]&=&S[\phi_1]~.
\eea
In order to find an expression for $\Gamma$ in terms of the classical field $\phi_c$, 
one needs to express the source $j$ in terms of $\phi_c$ , and plug the result in the expression (\ref{Gammasp}).
The classical field (\ref{phicsp}) can be written as
\bea
\phi_c&=&\frac{\phi_1+\phi_2}{2}-\frac{\phi_1-\phi_2}{2}\tanh\left(\frac{A_1-A_2}{2} \right)\nn
\mbox{where}~~A_i&=&S_i+\int j\phi_i~,~~~i=1,2~. 
\eea
As a consequence, since the fields $\phi_1,\phi_2,j$ are constant, we have
\be
\int j=\frac{S_1-S_2}{\phi_1-\phi_2}+\frac{2}{\phi_1-\phi_2}\tanh^{-1}\left( \frac{\phi_1+\phi_2-2\phi_c}{\phi_1-\phi_2}\right)~,
\ee
and therefore
\bea
A_i&=&\frac{\phi_1S_2-\phi_2S_1}{\phi_1-\phi_2}+\frac{2\phi_i}{\phi_1-\phi_2}\tanh^{-1}\left( \frac{\phi_1+\phi_2-2\phi_c}{\phi_1-\phi_2}\right)\nn
\int j\phi_c&=&\frac{\phi_c(S_2-S_1)}{\phi_1-\phi_2}+\frac{2\phi_c}{\phi_1-\phi_2}\tanh^{-1}\left( \frac{\phi_1+\phi_2-2\phi_c}{\phi_1-\phi_2}\right)~.
\eea
From these last two expressions, one finds that the effective action is
\bea
\Gamma[\phi_c]&=&\frac{(\phi_1-\phi_c)S_2+(\phi_c-\phi_2)S_1}{\phi_1-\phi_2}\nn
&&-\ln\left[\left( \frac{\phi_1-\phi_c}{\phi_c-\phi_2}\right)^\frac{\phi_c-\phi_1}{\phi_1-\phi_2} 
+\left(\frac{\phi_1-\phi_c}{\phi_c-\phi_2}\right)^\frac{\phi_c-\phi_2}{\phi_1-\phi_2}  \right] ~,
\eea
where the second term vanishes for $\phi_c=\phi_i$, $i=1,2$, and is independent of the volume $V$ of space time, 
whereas the first term is proportional to $V$. The expression for the effective potential 
$U_{eff}(\phi_c)=V^{-1}\Gamma[\phi_c]$ in terms of the the bare potential $U_{bare}(\phi)=V^{-1}S[\phi]$ 
is then obtained after dividing by $V$ and taking the limit $V\to\infty$:
\be
U_{eff}(\phi_c)=\frac{(\phi_1-\phi_c)U_{bare}(\phi_2)+(\phi_c-\phi_2)U_{bare}(\phi_1)}{\phi_1-\phi_2}~.
\ee
Therefore the effective potential is linear between $\phi_1$ and $\phi_2$ (see fig.1), 
and the concave part of the bare potential has been eliminated by the presence of non-trivial saddle 
points in the partition function (\ref{Z}). Note that the present argument is valid for an infinite volume.\\
If $S_1=S_2$, one recovers here the famous Maxwell construction, where the isothermal curve in the Clapeyron diagram shows a plateau,
corresponding to a constant saturated vapour pressure, as long as both vapour and liquid coexist.
In the quantum field theory case \cite{Miransky}, the vacuum $|0>$ is made of a superposition of the two states $|\phi_1>$ and $|\phi_2>$,
which satisfies $<0|\phi|0>=\phi_c$.

\begin{figure}
\label{Maxwell1} 
\epsfxsize=10cm
\centerline{\epsffile{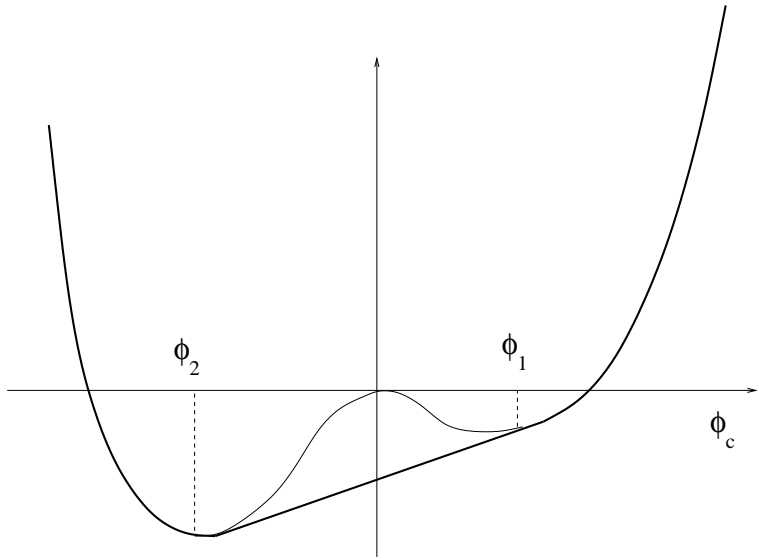}}
\caption{The effective potential (thick line) interpolates linearly the minima of the bare potential (thin line). Quantum corrections 
are neglected here, and would slightly modify the effective potential quantitatively only.}
\end{figure}

\subsection{Wilsonian approach}

The first studies \cite{Fukuda} and \cite{Tabata} show that non-trivial saddle point must be taken into account in Wilsonian 
renormalization procedure, for each blocking step in the construction of the infrared theory (IR).\\
Infinitesimal Wilsonian renormalization group studies have later also shown the flattening of the running potential in the IR. One way is to consider
the average action formalism (see for example \cite{Wetterich1} for a review), where a smoothly varying weight is associated to Fourier modes 
in the path integral.
IR modes, with momenta smaller than a typical scale $k$ are then frozen, and only UV modes, with momenta larger than $k$, are 
integrated out. 
As $k\to0$, the full effective action is recovered, and one can follow the evolution of the average action with $k$, in order 
to construct the full quantum theory in the limit $k\to0$. 
In \cite{Wetterich2}, a non-trivial saddle point dominates the path integral defining the average action, 
and as $k\to0$, the average potential smoothly goes to a flat potential in the IR.\\
A similar study is done in \cite{ABP}, using a sharp cut off, and shows that a non-trivial saddle point (plane wave with momentum $k$), 
in each infinitesimal integration of Fourier modes between $k-\delta k$ and $k$, leads to the flattening of the potential, 
as shown in fig.2. \\
The smoothness of renormalization flows in the presence of the so-called spinodal instability might not clearly be established yet.
Analytical argument for the absence of singularity in the renormalization flows are given in \cite{Litim}, but 
singular flows are observed in \cite{PolonyiPhi4}, where the study in based on precise numerical analysis.
It is interesting to note that the Sine-Gordon model must lead to a flat effective potential, for every value of the classical field, since
this is the only way for the effective potential to be convex and periodic at the same time. 
This has been studied in \cite{PolonyiSG}, where, unlike with a
polynomial bare potential, smooth renormalization flows are observed numerically. \\
Another example of flattening of a periodic bare potential
was looked at in \cite{AT} for the axion field, using an alternative to Wilsonian renormalization.

\begin{figure}
\label{Maxwell2} 
\epsfxsize=10cm
\centerline{\epsffile{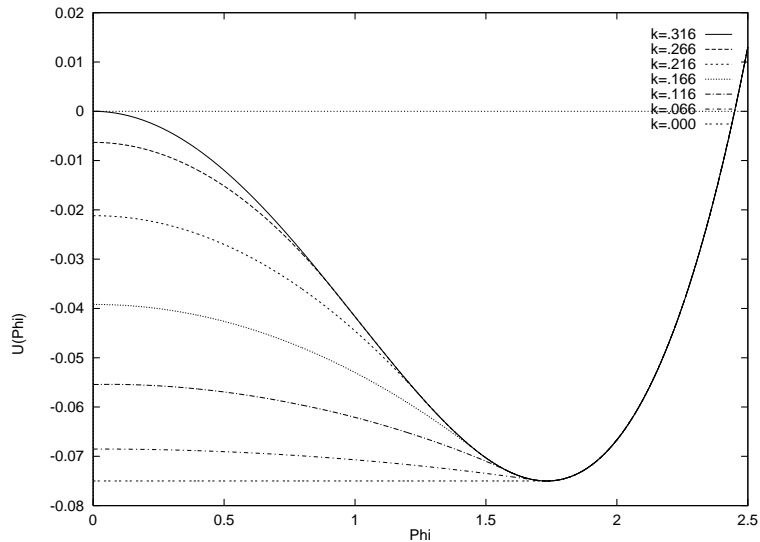}}
\caption{The Maxwell construction in infinitesimal Wilsonian renormalization group studies, involving the sharp cut off.
The running potential flatens as the observation scale reaches the IR (this figure is taken from \cite{ABP}). }
\end{figure}

\section{Yukawa model}

We add here to the scalar self interactions a Yukawa interaction, and show that the potential is also linear 
between the minima of the bare action.

\subsection{Convexity of the effective potential}

\nin The partition function, functional of the sources $j,\eta,\ol\eta$, is
\be
Z[j, \eta, \ol\eta]=\int{\cal D}[\phi,\psi,\ol\psi]\exp\left(-S[\phi,\psi,\ol\psi]
-\int j\phi+\ol\eta\psi+\ol\psi\eta\right)\equiv\exp\left( -W[j, \eta, \ol\eta]\right) ~,
\ee
and the classical fields are defined as
\be
\phi_c=\frac{\delta W}{\delta j}=\left<\phi\right>~~,~~
\psi_c=\frac{\delta W}{\delta\ol\eta}=\left<\psi\right>~~,~~
\ol\psi_c=-\frac{\delta W}{\delta\eta}=\left<\ol\psi\right>\nonumber~.
\ee
The effective action is
\be
\Gamma[\phi_c,\psi_c,\ol\psi_c]=W[j,\eta,\ol\eta]-\int \left( j\phi_c+\ol\eta\psi_c+\ol\psi_c\eta\right) ~,
\ee
where the sources $j,\eta,\ol\eta$ must be understood as functionals of the classical fields. The
equation of motion for the classical scalar field is then
\bea
\frac{\delta\Gamma}{\delta\phi_c}&=&\int\frac{\delta W}{\delta j}\frac{\delta j}{\delta\phi_c}
-\frac{\delta W}{\delta\eta}\frac{\delta\eta}{\delta\phi_c}+\frac{\delta\ol\eta}{\delta\phi_c}\frac{\delta W}{\delta\ol\eta}
-j-\int\left(\frac{\delta j}{\delta\phi_c}\phi_c+\frac{\delta\ol\eta}{\delta\phi_c}\psi_c+\ol\psi_c\frac{\delta\eta}{\delta\phi_c}\right)\nn
&=&-j ~,
\eea
and, similarly, we find
\be
\frac{\delta\Gamma}{\delta\psi_c}=\ol\eta\nn~~~~\mbox{and}~~~~\frac{\delta\Gamma}{\delta\ol\psi_c}=-\eta~,
\ee
such that the second derivative matrices $\delta^2\Gamma$ and $\delta^2 W$ satisfy
\be\label{inverse}
\delta^2\Gamma=-(\delta^2 W)^{-1}~.
\ee
The off-diagonal terms in $\delta^2\Gamma$ vanish when setting 
the fields to the constant values $\phi_c=\phi_0$, $\ol\psi_c=\psi_c=0$:
\be
\left.\frac{\delta^2\Gamma}{\delta\phi_c\delta\psi_c}\right|_0
=\left.\frac{\delta^2\Gamma}{\delta\phi_c\delta\ol\psi_c}\right|_0
=\left.\frac{\delta^2\Gamma}{\delta\psi\delta\psi_c}\right|_{vac}
=\left.\frac{\delta^2\Gamma}{\delta\ol\psi_c\delta\ol\psi_c}\right|_0=0~,
\ee
such that
\be
\delta^2\Gamma_0=
\left(\begin{array}{ccc}
\delta^2\Gamma_{\phi_c\phi_c}&0&0\\
0&\delta^2\Gamma_{\psi_c\ol\psi_c}&0\\
0&0&\delta^2\Gamma_{\ol\psi_c\psi_c}
\end{array}\right)~.
\ee
As a consequence, because of the inequality (\ref{1stkey}),
\be
(\delta^2\Gamma_{\phi_c\phi_c})^{-1}=(\delta^2\Gamma^{-1})_{\phi_c\phi_c}=-(\delta^2 W)_{jj}\ge0~,
\ee 
and the scalar sector is convex, as in the self-interacting case. But the additional feature here is the 
possibility of having a fermion condensate, which contributes to the effective potential for the scalar field, as we show
in the next section.

\subsection{Linear effective potential}

We consider here the following Euclidean model
\be
S[\phi,\ol\psi,\psi]=\int d^4x\left(\frac{1}{2}\partial_\mu\phi\partial^\mu\phi+\ol\psi i\br\partial\psi+g\phi\ol\psi\psi +U_{bare}(\phi)\right)~,
\ee
where the potential $U_{bare}$ features SSB characteristics. For the sake of simplicity, we assume that the two 
minima are located at opposite points $\pm\phi_0$ and have the same value $U_{bare}(\pm\phi_0)\equiv U_0$.
In what follows, we consider constant fermionic configurations: the effective scalar potential will depend on the fermion condensate only,
which is a scalar and can have a non-vanishing uniform v.e.v.\\  
Considering the same approximation as in section 2, the partition function is then
\bea
Z&\simeq&\exp\left( -\int U_0+j\phi_0\right)\int{\cal D}[\ol\psi,\psi]\exp\left( -\int g\phi_0\ol\psi\psi+\ol\eta\psi+\ol\psi\eta\right)\nn
&&+  \exp\left( -\int U_0-j\phi_0\right)\int{\cal D}[\ol\psi,\psi]\exp\left( -\int -g\phi_0\ol\psi\psi+\ol\eta\psi+\ol\psi\eta\right)~,
\eea
We define then
\bea
\chi_+=\psi+\frac{\eta}{g\phi_0}~~,~~~~\chi_-=\psi-\frac{\eta}{g\phi_0}~,
\eea
to obtain
\bea
Z&\simeq&\exp\left( -\int U_0+j\phi_0\right)\int{\cal D}[\ol\chi_+,\chi_+]
\exp\left( -\int g\phi_0\ol\chi_+\chi_+-\frac{\ol\eta\eta}{g\phi_0}\right)\nn
&&+  \exp\left( -\int U_0-j\phi_0\right)\int{\cal D}[\ol\chi_-,\chi_-]
\exp\left( \int g\phi_0\ol\chi_-\chi_--\frac{\ol\eta\eta}{g\phi_0}\right)~.
\eea
The integration over fermions is dominated by a possible fermion condensate $\rho=<\ol\psi\psi>$, such that
\bea
Z&\simeq&\exp\left(-\int U_0+j\phi_0+g\phi_0\rho-\frac{\ol\eta\eta}{g\phi_0} \right) 
+\exp\left(-\int U_0-j\phi_0-g\phi_0\rho+\frac{\ol\eta\eta}{g\phi_0} \right)\nn
&=&2\exp\left( -\int U_0\right)  \cosh(A)~,
\eea
where 
\be\label{Abis}
A=\int \left( j\phi_0+g\phi_0\rho-\frac{\ol\eta\eta}{g\phi_0}\right) ~.
\ee
There is actually a zero mode which should be taken into account in the evaluation of the dominant contributions for $Z$, 
due to the global symmetry $\psi\to e^{i\theta}\psi$ and $\ol\psi\to e^{-i\theta}\ol\psi$, but the corresponding
factor is source-independent and has no consequence on the results. \\
The classical fields are
\bea
\phi_c&=&-\frac{1}{Z}\frac{\delta Z}{\delta j}=-\phi_0\tanh(A)\nn
\psi_c&=&-\frac{1}{Z}\frac{\delta Z}{\delta\ol\eta}=\frac{\eta}{g\phi_0}\tanh(A)\nn
\ol\psi_c&=&~~\frac{1}{Z}\frac{\delta Z}{\delta\eta}=\frac{\ol\eta}{g\phi_0}\tanh(A)~,
\eea
and we see that the present approximation is valid only in the interval $[-\phi_0,\phi_0]$, since the limits of $\phi_c$ are
\be
\lim_{j\to+\infty}\phi_c=-\phi_0~~,~~\mbox{and}~~~~\lim_{j\to-\infty}\phi_c=\phi_0~.
\ee
The condensate $\rho$ is defined 
for vanishing fermionic sources, and with the classical scalar field set to its v.e.v
$v$ (the corresponding source is $j(v)$):
\bea
\rho&\equiv&\frac{1}{Z}\int{\cal D}[\phi,\ol\psi,\psi]~\ol\psi\psi\exp\left(-S[\phi,\psi,\ol\psi]-\int j(v)\phi\right)\nn
&=&\left(\frac{\delta^2W}{\delta\eta\delta\ol\eta} \right)_{\eta=\ol\eta=0,\phi_c=v} \nn
&=&-\left(\frac{1}{Z}\frac{\delta^2Z}{\delta\eta\delta\ol\eta} \right)_{\eta=\ol\eta=0,\phi_c=v}\nn
&=&\left( \frac{\tanh(A)}{g\phi_0}\delta(0)\right)_{\eta=\ol\eta=0,\phi_c=v}~,
\eea
where $\delta(0)=\lim_{x\to0}\delta(x)$ is the volume of Fourier space, that we denote $\Lambda^4$, where $\Lambda$ is an ultraviolet (UV)
cut off.
The latter identity is a self consistent equation which determines $\rho$, but one can express the fermion condensate in terms of the 
scalar field v.e.v 
\be
v=[-\phi_0\tanh(A)]_{v.e.v}~,
\ee
to obtain finally
\be
\rho=-\frac{\Lambda^4v}{g\phi_0^2}~.
\ee
Using these different results, and expressing $j$ in terms of $\phi_c$ as
\be
\left( \int j\phi_0\right)_{\ol\eta=\eta=0} =-\tanh^{-1}\left( \frac{\phi_c}{\phi_0}\right)-\int g\phi_0\rho~, 
\ee
we find that the scalar sector of the effective action is, for $\phi_c=$ constant,
\be
\Gamma_{scalar}=\int (U_0+g\phi_c\rho)
-\ln\left( 2\cosh\tanh^{-1}\left(\frac{\phi_c}{\phi_0}\right)\right)  
+\frac{\phi_c}{\phi_0}\tanh^{-1}\left( \frac{\phi_c}{\phi_0}\right)~.
\ee
The last two terms in the previous equation are both divergent when $\phi_c\to\pm\phi_0$, but these divergences cancel,
such that together these terms give a finite contribution, independent of the volume $V$ of space time. 
The effective potential is finally obtained after dividing by $V$ and taking the limit $V\to\infty$, where
the only remaining terms are 
\be
U_{eff}(\phi_c)= U_0+g\phi_c\rho~.
\ee
The effective potential is therefore linear between the minima of the bare potential. We represent $U_{eff}(\phi_c)$ in fig.3, where
the smooth matching with the bare potential outside the interval $[-\phi_0,\phi_0]$ is assumed, and arises from quantum corrections, which are not 
taken into account in this study.\\
As can be seen on fig.3, the minimum of the effective potential occurs for the v.e.v $v=\phi_0$, 
since the condensate is negative and is given by
\be\label{condensate}
\rho=-\frac{\Lambda^4}{g\phi_0}~.
\ee
Note that the latter result does not make sens for $\phi_0=0$, but the whole study is
valid for a non-vanishing interval $[-\phi_0,\phi_0]$ only.

\begin{figure}
\label{Maxwell3} 
\epsfxsize=10cm
\centerline{\epsffile{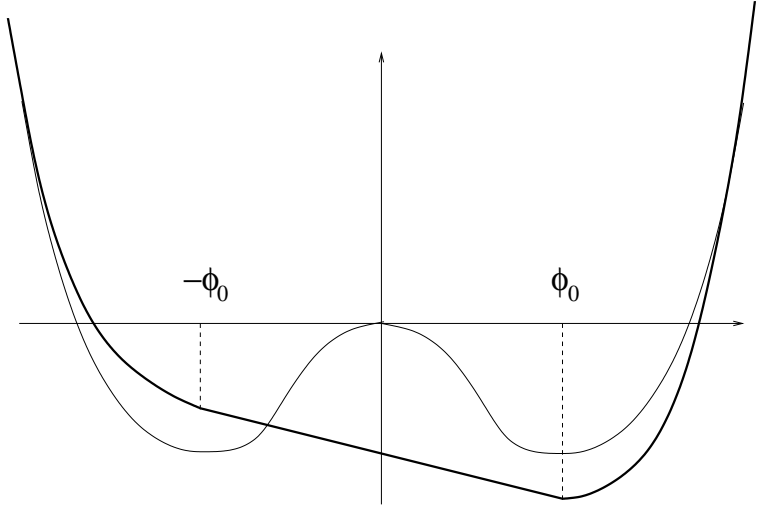}}
\caption{The effective potential is linear in the interval $[-\phi_0,\phi_0]$, with a slope due to the fermion condensate.
The smooth matching with the bare potential outside the interval $[-\phi_0,\phi_0]$ is assumed, and arises from quantum corrections,
which are not evaluated here. }
\end{figure}

\section{Conclusion}

We showed with a semi classical approach that the flattening of the effective potential can be obtained after approximating
the partition function by its main contributions, arising from non-trivial saddle points.
In the case of the Yukawa model, it is interesting to see that the fermion condensate and the scalar v.e.v are self-consistent,
and arise from a linear effective potential. The situation where the fermion condensate vanishes is equivalent to the 
absence of a specific scalar v.e.v, and is consistently represented by a flat effective potential. \\
The next step is to consider a complex scalar field, and to couple it to a gauge field. This case is more subtle though, 
because of the existence of zero modes and non-trivial manifold of saddle points in the 
partition function.
Such a study has been initialized in \cite{Hindmarsh}, using the interpolated loop expansion, as well as
a t'Hooft-style gauge fixing, which involves the Higgs field and reduces the manifold of saddle points. The Authors show that 
the scalar effective potential can be made convex to any order of the loop expansion. We plan to extend the present work to such a situation,
in order to find non-perturbative constraints on the Higgs potential.

\end{document}